\documentclass[astrosymb, twocolumn]{aastex701}

\usepackage{newtxtext}
\usepackage{newtxmath}

\usepackage{amsmath}	% Advanced maths commands
\usepackage{amssymb}	% Extra maths symbols
\usepackage{mathrsfs}   % Extra extra math symbols
\usepackage{enumitem}
\usepackage{verbatim} %for comment command
\usepackage[caption=false]{subfig}
\usepackage{array}
\usepackage{float}
\usepackage{listings}
\usepackage{hyperref}
\usepackage{blindtext}
\usepackage{physics}
\usepackage{gensymb}

\usepackage{anyfontsize}
\usepackage[T1]{fontenc}
\usepackage{etoolbox}

\begin{document}

\title[]{Radio prospects of extrasolar aurorae polaris as a probe of planetary magnetism}
% Remove text indicating that this is a draft 
%\makeatletter\let\frontmatter@title@above=\relax 

% command to be used to indicate a to-do item
\newcommand{\todo}[1]{\textcolor{red}{ToDo: #1}}

\newcommand{\done}[1]{\textcolor{teal}{Done: #1}}

%command to be used to indicate a point that is interesting or not clearly understood
\newcommand{\hmm}[1]{\textcolor{yellow}{Hmm: #1}}

%Insert text in math equations
\newcommand{\mt}[1]{\mathrm{#1}}

\newcommand{\ffc}{\mbox{55 Cnc e}}
\newcommand{\wb}{\mbox{WASP-18 b}}
\newcommand{\taub}{\mbox{tau Boo b}}

%\email{asaf.kaya@ug.bilkent.edu.tr}
\author[0009-0001-6374-2728]{Asaf Kaya}
\altaffiliation{Email: \href{mailto:asaf.kaya@ug.bilkent.edu.tr}{asaf.kaya@ug.bilkent.edu.tr}}
\affiliation{Department of Physics, Bilkent University, Bilkent 06800, Ankara, Türkiye}
\email{asaf.kaya@ug.bilkent.edu.tr}

\author[0000-0002-6939-9211]{Tansu Daylan}
\altaffiliation{Email: \href{mailto:tansu@wustl.edu}{tansu@wustl.edu}}
\affiliation{Department of Physics and McDonnell Center for the Space Sciences, Washington University, St. Louis, MO 63130, USA}
\email{tansu@wustl.edu}

%\correspondingauthor{Tansu Daylan}

\shorttitle{Detectability of Exoplanetary Radio Emissions}
\shortauthors{A. Kaya and T. Daylan}

% Please follow my suggested checklist when working on the paper draft:
% https://drive.google.com/file/d/13Usk-OFk8G_34zu93B2AKq3k_XDUa-_j/view?usp=share_link

\begin{abstract}
Magnetized exoplanets are expected to emit auroral cyclotron radiation in the radio regime due to the interactions between their magnetospheres, the interplanetary magnetic field, and the stellar wind. Prospective extrasolar auroral emission detections will constrain the magnetic properties of exoplanets, allowing the assessment of the planets' habitability and their protection against atmospheric escape by photoevaporation, enhancing our understanding of exoplanet formation and demographics. We construct a numerical model to update the estimates of radio emission characteristics of the confirmed exoplanets while quantifying the uncertainties of our predictions for each system by implementing a Monte Carlo error propagation method. We identify \added{16} candidates that have expected emission characteristics that render them potentially detectable from current ground-based telescopes. Among these, the hot Jupiter \mbox{tau Boötis b} \added{is the most favorable target with an} expected flux density \added{of} $51^{+36}_{-22}$ mJy. Notably, \added{eleven} candidates are super-Earths and sub-Neptunes, for which magnetism is key to understanding the associated demographics. Together with the other predictive works in the literature regarding the characteristics and the geometry of the magnetospheric emissions, our predictions are expected to guide observational campaigns in pursuit of discovering magnetism on exoplanets.
\end{abstract} 

%Keywords function was modified in the cls file, along with the post abstract space.

\section{Introduction}
\label{section: Introduction}
The observed bimodal radius distribution of close-in, small exoplanets \citep{Owen2013} and the underabundance of highly irradiated Neptune-sized planets \citep{Szabo2011, Lundkvist2016, Owen2018} suggest that exoplanets can lose a significant fraction of their volatile envelopes via atmospheric loss \citep{Lopez2013, Jin2014, Fulton2017, Owen2017, Eylen2018, Gupta2019, Venturini2020}. The recent accelerated growth in the number of multiplanetary systems amenable to comparative atmospheric characterization \citep[e.g., ][]{Guenther2019, Badenas-Agusti2020, Guerrero2021, Daylan2021} identified by the Transiting Exoplanet Survey Satellite \citep[TESS; ][]{Ricker2015} present an opportunity to address our knowledge gaps in the atmospheric retention of small exoplanets \citep{Owen2019b, Carolan2021, Luo2023, Chin2024}.

Despite the progress in the characterization and modeling of the demographics of small exoplanets \citep{Lopez2014, Chen2016, Ginzburg2018, Owen2019a}, an outstanding observational knowledge gap remains regarding the level of magnetism present in exoplanets subject to atmospheric loss as well as whether the magnetization of a planet supports or impedes its ability to retain an atmosphere \citep{Owen2014, Gronoff2020}. A strong and stable magnetic field can slow down the loss of the volatile envelope due to ionization caused by high energy radiation as well as Coronal Mass Ejections (CME) \citep{Khodachenko2007, Adams2011, Trammell2011, Owen2014, Trammell2014, Khodachenko2015, Green2021}. It can also reinforce the habitability of a planet by protecting its surface against ionizing radiation and cosmic rays \citep{Grießmeier2005b, Grießmeier2009b, Dartnell2011, Griesßmeier2015}. However, the magnetization of a planet can also facilitate ion escape from magnetic poles \citep{Gunell2018, Egan2019}. Either way, planetary magnetism plays a major role in sculpting the observed exoplanet demographics, and magnetic characterization of exoplanets stands as a compelling frontier in advancing our understanding of their atmospheric loss mechanisms as well as characterizing the habitability prospects beyond simple considerations based on their irradiation state \citep{Lammer2007, Lammer2009, Grießmeier2009a}. 

While there has not yet been a confirmed detection of exoplanetary magnetism \citep{Lazio2018, Shiohira2024}, several strategies have been put forward over the last decades. Two promising approaches are the enhancement of the host star's chromospheric activity due to the magnetized star‐planet interactions (SPI) with close‐in exoplanets and the associated optical signatures \citep{Cuntz2000, Shkolnik2003, Shkolnik2005, Shkolnik2008, Cauley2019} and the ingress-egress asymmetries in the near‐UV light curves of exoplanets within the presence of bow shocks \citep{Vidotto2010, Vidotto2011, Llama2011}. Recently, to constrain the magnetic fields of exoplanets, \cite{Schreyer2024} and \cite{Savel2024} proposed spectral analysis of exoplanet atmospheres, the former utilizing the 1083 nm Helium line and the latter measuring the differences between the velocities of neutral gas and heavy ions in them since magnetic fields directly deflect ions. These two approaches require analyses of high-resolution spectroscopy data in the near-IR, visible, and near-UV bands.

On the other hand, auroral radio emissions associated with the magnetized exoplanet is a promising probe of exoplanetary magnetic fields \citep{Winglee1986, Zarka1997, Farrell1999}. Historically, the decametric radio emission from the planet Jupiter was the first to be detected \citep{Burke1955}. Not shortly after, it was understood that this emission was linked to the planet's magnetic field \citep{Carr1969}, and other planets such as Earth and Saturn had similar radio emissions \citep{Gurnett1974, Kaiser1980}. With these revolutionary observational developments and the seminal theoretical work of \cite{Wu1979} establishing the Cyclotron Maser Instability (CMI) mechanism behind the observed emissions, a new domain of detecting otherwise quiet celestial objects—radio—emerged and thus began a new era in planetary science. Since then, understanding, detecting, and interpreting the prospective radio emissions of exoplanets has been a significant problem in astronomy \citep{Charbonneau2004, Farrell2004, Butler2005, Grießmeier2006b, Lazio2009, Grießmeier2015b}.

The motivation for the search of these radio signals is manifold. In addition to the demographics and habitability considerations, the magnetic field strengths sought to be determined from the frequency and intensity of such signals will constrain the models of the internal structure of planets \citep{Sanchez2004} and enable testing empirical scaling laws on their magnetic fields \citep{Farrell1999}. Further, the prospective measurement of the dynamic spectra of the intensity and polarization of the radio emissions from exoplanets will be imperative for the field. \cite{Hess2011} showed using exhaustive simulations that such measurements could divulge some of the vital parameters of exoplanetary systems—in addition to the strength and tilt of the planet's magnetic field, the orbital inclination, and the rotation and revolution periods. Thus, a dynamic analysis of these radio signals would not only shed light on SPIs but also eliminate the ambiguity on the planet's mass by constraining the inclination. Moreover, it would effectively test the tidal synchronization theories with the inference of the rotation periods. Finally, in exoplanet systems having induced emissions via a satellite (Jupiter-Io analog), an opportunity to detect exomoons is born.

While it is natural, extrapolating from the solar system, to expect that exoplanets should emit in radio wavelengths, their emissions must extend over a much larger range compared to those observed in the solar system. Studies have been conducted to predict and evaluate the observability of such emissions \citep{Winglee1986, Zarka2001, Farrell2003, Lazio2004, Grießmeier2005a, Griessmeier2007a, Reiners2010, Zarka2015a, Lynch2018, Turnpenney2018, Vidotto2019, Zarka2019, Ashtari2022, Li2023, Mauduit2023a}. However, since the closest exoplanets are $\sim10^5$ times further away than the furthest solar system planet from Earth, their radio signals are at least $10^{10}$ times weaker than those of the planets in the solar system. This drastic decrease in the intensity of emissions renders it challenging to observe exoplanets in the radio domain. Despite decades of observational campaigns that have been carried out \citep{Yantis1977, Winglee1986, Zarka1997, Lazio2007, Lynch2017, GreenDA2021} and a few cases of tentative detections \citep[e.g., ][]{Etangs2013, Sirothia2014, Turner2021}, no exoplanetary radio emission have so far been confirmed \citep{Bastian2000, Lazio2004, George2007, Smith2009, Stroe2012, Hallinan2013, Murphy2015, O'Gorman2018, Lenc2018, Lynch2018, deGasperin2020, Narang2021,Turner2024}.

In this work, we investigate the direct detection prospects of auroral radio emissions from magnetized exoplanets due to CMI, developing a computational framework to predict the radio flux densities and frequencies of characteristic magnetospheric emissions from confirmed exoplanets. We uniquely employ a Monte Carlo error propagation method in the calculations to assess the associated uncertainties. Considering the limitations of the current observing facilities, we then evaluate the observability of the auroral emissions of the exoplanets in our sample. We maintain our model in a publicly available software pipeline, \emph{Aegis}\footnote{\url{https://github.com/AstroMusers/aegis}}
, with an extensive library for the research community to reproduce and potentially expand the framework \citep{aegis}. 
\added{Notably, our model does not take into account the eclipsing of the expected planetary radio emissions by the wind of the host star, through the process of free-free absorption \citep{Kavanagh2020}. While detailed modeling of this effect is necessary, and might be possible for individual systems \citep[e.g.,][]{Kavanagh2023}, it is currently impractical for a large sample of targets such as the one analyzed in this work.}

The rest of this paper is structured as follows. In Section~\ref{section: Model}, we discuss the model we use to estimate the radio luminosity and emission frequencies of exoplanets. In Section~\ref{section: Methodology}, we explain how we determine the largest suitable exoplanet sample, along with our methods to handle the parameters of the model to propagate the associated initial uncertainties. In Section~\ref{section: Results}, we present our predictions for radio emissions from exoplanets in our sample, evaluating observability prospects by their emission characteristics and considering their relative position in the sky. In Section~\ref{section: Discussion}, we discuss our results, compare them to previous predictions and observations, and evaluate consistencies. We also discuss the possible implications of prospective detections\added{, explain the theoretical limitations of our model,} and provide an outlook. We conclude in Section~\ref{section: Conclusion} with a summary.

\section{Planetary Radio Emission Model}
\label{section: Model}

\subsection{Electron Cyclotron Maser Instability (CMI)}
\label{subsection: CMI}
Electrons in a planetary magnetosphere can be accelerated to higher energies through various processes, including the stellar wind kinetic and magnetic energies, coronal mass ejections, magnetospheric plasma sources, and unipolar interaction (for a review, see \cite{Zarka2018}). These accelerated electrons move along the magnetic field lines of the planet's magnetic field, gyrating around the lines and producing circularly polarized cyclotron radiation. Cyclotron radiation with a frequency below the characteristic plasma frequency is prevented from propagating and escaping from the ambient plasma of the magnetosphere since emissions with such frequencies are absorbed by the plasma. However, radiation with a high enough frequency can escape. Several effects combine to allow the escape of intense cyclotron radiation from the magnetosphere, including the relativistic Doppler shift of extremely high-speed electrons. In addition, if the electron energy distribution consists of more electrons that can escape from the ambient plasma than those that cannot, the system is fed to produce even more intense, coherent radiation through the mechanism called the Electron Cyclotron Maser Instability \citep{Wu1979, Wu1985}. In most cases, CMI is the leading mechanism that drives the planets to emit in radio wavelengths \citep{Zarka2001, Treumann2006}.

\subsection{CMI‐Driven Emissions Caused by Stellar Wind Interactions}
\label{subsection: CMI-stellar}

Radiometric Bode's Law \citep[RBL; ][]{Desch1984} states that the radio output power of an exoplanet emitting with CMI is proportional to the incident kinetic and magnetic power on the exoplanet provided by the stellar wind of its host star,
\begin{equation}
    P_\mt{rad} = \epsilon P_\mt{in},
    \label{RBL}
\end{equation}
where $\epsilon$ is a proportionality constant, and $P_\mt{rad}$ and $P_\mt{in}$ parameterize the output radio power and the incident power, respectively.

Generally, all of the previously mentioned energy sources presented in \cite{Zarka2018} (up to keV energy levels) may cause auroral cyclotron emission in a planet's magnetosphere driven by CMI. Leaving out the less predictable nature of coronal mass ejections and internal plasma sources and deeming unipolar interaction out of the scope of this work, we focus on the interaction between the stellar wind and the planetary magnetosphere. In this case, one could consider two major input energy sources. The first is the kinetic energy supplied by the stellar wind, initially considered by \cite{Desch1984}. Assuming the stellar wind is made up of protons, the incident kinetic power relation is
\begin{equation}
    P_\mt{in, kin} = m_p n v_\mt{eff}^3 \pi R_\mt{mp}^2,
    \label{inkin}
\end{equation}
where $m_p$ is the proton mass, $n$ is the number density of wind particles, $v_\mt{eff}$ is the effective speed of the planet in the stellar wind, and $R_\mt{mp}$ is the magnetopause standoff distance associated with the planet's magnetosphere. 
The second energy source is the magnetic energy supplied by the wind, first proposed by \cite{Zarka2001}. The incident magnetic power relation is, considering the Poynting flux on the magnetospheric cross-section of the planet induced by the magnetized wind,
\begin{equation}
    P_\mt{in, mag} = \frac{B_\perp^2}{2\mu_0} v_\mt{eff} \pi R_\mt{mp}^2,
    \label{inmag}
\end{equation}
where $B_\perp$ is the component of the interplanetary magnetic field (IMF) perpendicular to the stellar wind flow, and $\mu_0$ is the vacuum permeability constant. 
Along the lines of \cite{Griessmeier2007b}, we scale Equation~\ref{RBL} with Jovian radio emission. We take the average high activity emission power $P_\mt{rad, J} = 2.1 \times 10^{18}\,\mt{\added{erg}}\,\mt{\added{s}}^{\added{-1}}$ \citep{Zarka2004}, and compute $P_\mt{in, J}$ both from the incident kinetic power and magnetic power to find two proportionality constants $\epsilon_\mt{mag} = 6.4 \times 10^{-5}$, $\epsilon_\mt{kin} = 1.5 \times 10^{-6}$. The interpretation of the proportionality constants is as the efficiencies of the different energy sources to accelerate the electrons in the planet's magnetosphere, leading to CMI \citep{Desch1984, Zarka2001}. To account for both of the energy sources without changing the relative efficiencies of different mechanisms, assuming the two mechanisms are independent, the only choice of $P_\mt{rad}$ congruent with the RBL is 
\begin{equation}
    P_\mt{rad} = \frac{1}{2}(\epsilon_\mt{kin} P_\mt{in, kin} + \epsilon_\mt{mag} P_\mt{in, mag}),
\end{equation}
which determines the main radio output power of exoplanets. While many works consider these mechanisms separately or focus solely on the magnetic power transmission, such a choice of $P_\mt{rad}$ is unique in the literature.

\subsection{Emission Characteristics}
According to the CMI theory, a planet will radiate up to a specific maximum frequency, where the conditions for CMI apply \citep{Farrell1999}. Practically, these are the regions near the magnetic poles. Therefore, CMI-driven radio emissions have a clear cutoff, given by the cyclotron resonance frequency
\begin{equation}
    \begin{split}
        \nu_{\mathrm{max}} & =  \frac{e\mathscr{M}R^3}{2\pi m_e} =\frac{e B_{\mathrm{pole}}}{2\pi m_e} = 2.8\mathrm{MHz} \left( \frac{B_{\mathrm{pole}}}{\mathrm{1G}} \right)
    \end{split},
    \label{maxnu}
\end{equation}
where $e$ is the unit charge, $\mathscr{M}$ is the magnetic moment, $R$ is the planetary radius, $m_e$ is the mass of the electron, and $B_\mt{pole}$ is the magnetic field strength in the cloud tops in the magnetic polar regions, which is assumed to be relevant for CMI. 
The propagation condition for CMI emission is then $\nu_\mt{max} > \nu_p$ where $\nu_p$ is the characteristic plasma frequency of the medium and is given as
\begin{equation}
    \nu_p = \sqrt{\frac{ne^2}{\pi m_e}} = 8.98 \, \mt{ kHz} \, \sqrt{n},
    \label{eq: plasmafreq}
\end{equation}
where $e$ and $m_e$ are the electronic charge and mass, respectively, and $n$ is expressed in $\mt{cm}^{-3}$. Since the particle density of the wind decreases with the distance from the host star, this relation is generally most restrictive at the planet's position. However, depending on the orbital inclination of the system, the emission might have to propagate through denser wind regions of the stellar system on its way to Earth. Given the reliance of most exoplanet discovery methods on $\sim 90\degree$ inclinations, this problem becomes a considerable one for different orbital phases of the planet at times of observation.

Consistent with Equation~\ref{maxnu}, Jupiter's CMI-driven radio emission has a cutoff at around 40 MHz, and the Jovian magnetic field strength at the cloud tops is around 14 G \citep{Lazio2018}. This is also consistent with the analysis conducted by \cite{Connerney2018} on Juno data that shows that the Jovian surface magnetic field strength ranges from 2 to 20 Gauss in different locations, generally reaching higher values near polar regions. 

Assuming the spatial orientation of the exoplanet and the Earth allows the beamed CMI emissions to reach Earth, the radio emission flux density, which is assumed to be constant in the emission spectrum consistent with Jupiter's emission \citep{Zarka2004a}, measured from the Earth is \citep{Farrell1999}:
\begin{equation}
    \Phi = \frac{P_\mt{rad}}{\Omega D_\star^2 \Delta \nu} \, [10^{26} \, \mt{Jy}],
    \label{eq: flux}
\end{equation}
where $D_\star$ is the distance to the exoplanetary system, $\Delta \nu$ is the bandwidth of the emission taken to be the frequency at maximum intensity, and $\Omega$ is the solid angle of the emission that accounts for the beamed nature of the signal. In their analysis of the Jovian radio emissions and their spectra using Cassini observations, \cite{Zarka2004} report $\Omega = 1.6 \, \mt{sr}$, and it has become standard practice to use this value as the beaming constant for exoplanets. 

\begin{figure}
    \centering
    \includegraphics[width=0.475\textwidth]{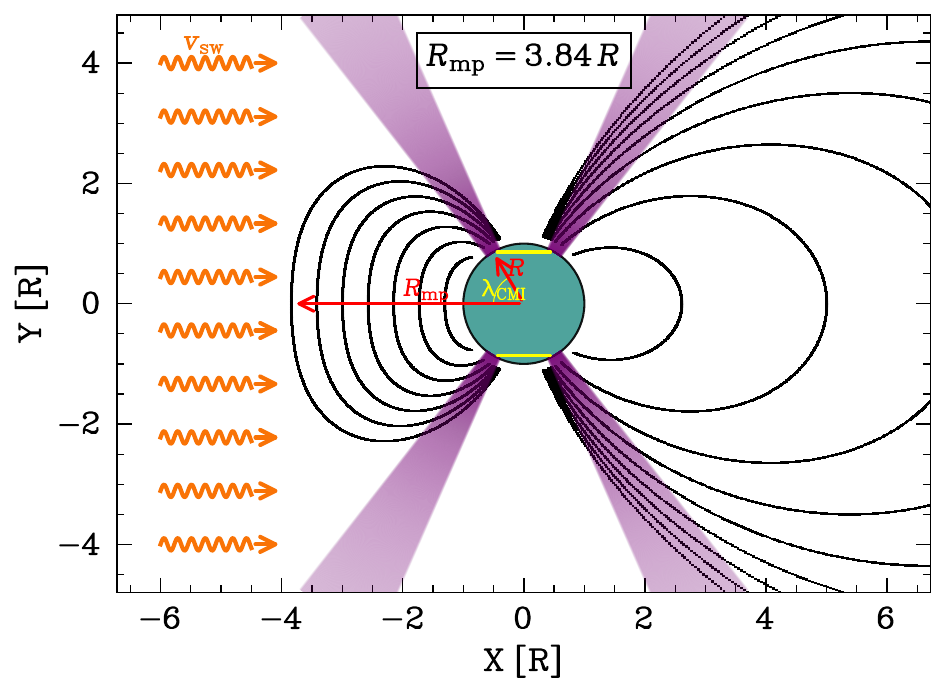}
    \caption{Schematic drawing of a magnetized planet experiencing pressure from the stellar wind flowing from the left side. The black magnetic field lines are drawn simply to indicate the dayside compression and nightside elongation of the magnetosphere. Therefore, they are only representative and not calculated using MHD simulations. The magnetopause standoff distance and the associated latitude of emission are shown, along with the purple, hollow cone of emission.}
    \label{fig: geometry_emission}
\end{figure}

Further, the existence of auroral radio emissions does not guarantee their visibility from Earth. A reasonable approximation for the emission latitude of the CMI-driven emissions is \citep{Ashtari2022}:
\begin{equation}
    \lambda_{\mt{CMI}} = \arccos{\left(\sqrt{\frac{R}{R_\mt{mp}}}\right)},
\end{equation}
where $R$ is the planetary radius and $R_\mt{mp}$ is the magnetopause standoff distance. In this convention, $\lambda \sim 90 \degree$ refers to polar regions, whereas $\lambda \sim 0 \degree$ corresponds to equatorial regions. Since the radiation is emitted almost entirely at the proximity of this latitude, it may not be visible from Earth due to the spatial orientation of the target system and the Earth. A visualization of the local geometry of the problem near the target planet, incorporating the emission latitudes and the beamed emission cone of a solid angle $\Omega = 1.6\,\mt{sr}$, is given in Figure~\ref{fig: geometry_emission}.
We assumed that the radiation is emitted from a latitude band centered around $\lambda_\mt{CMI}$ and calculated the width of this band so that its total solid angle with both hemispheres included is $1.6\, \mt{sr}$. Although Figure~\ref{fig: geometry_emission} accurately captures the anisotropy and the latitude dependence of the emission, the direction and the shape of the hollow cone are only meant to offer a fiducial illustration since they depend on the shape of the magnetic field lines.

After these considerations, Equation~\ref{eq: flux} is the ultimate equation to predict the radio emission flux densities of exoplanets at Earth distance, wherein the beamed nature of the emission is also accounted for through the assumption of the constant ($< 4\pi$) solid angle of emission. \cite{Zarka2004} further showed that the emission power $P_\mt{rad}$ fluctuates within different timescales, reaching up to emission powers about an order of magnitude higher than the average high activity emission power. Therefore, we determine the flux density of such burst emissions, $\Phi_\mt{peak}$, by multiplying the one obtained from Equation~\ref{eq: flux} by a constant factor of 10.

\subsection{The Stellar Wind Model}
\label{subsection: wind model}

As apparent from the dependencies of Equations~\ref{inkin} and~\ref{inmag}, the stellar wind properties significantly influence the results of the radio power obtained through RBL. Discussed further in Section~\ref{subsection: P. Properties}, the magnetopause standoff distance is determined by a pressure balance between external energy sources originating from the stellar wind and the internal magnetic field of the planet near the magnetosphere boundary. Hence, a treatment of stellar wind is inevitable.

In their analysis of different stellar wind conditions, \cite{Griessmeier2007a} show that the Parker solar wind model adequately describes the radial behavior of the stellar wind for stars with ages $\geq 0.7 \, \mt{Gyr}$. Further, \cite{Lynch2018} quantified the differences in radial wind profiles of younger stars computed from the Parker model to three-dimensional magnetohydrodynamic (MHD) simulations and found them to be small enough to be suitable for the order of magnitude calculations of the radio emission flux densities of exoplanets such as the one attempted in this work. Therefore, to determine wind speeds near the exoplanets of interest, we take the isothermal wind solution of \cite{Parker1958},
\begin{equation}
    \frac{v(r)^2}{c_s^2} - \ln{\left(\frac{v(r)^2}{c_s^2}\right)} = 4\ln{\left(\frac{r}{r_c}\right)} + 4\frac{r_c}{r} -3.
    \label{ParkerWind}
\end{equation}
Here, $r_c$ is the critical radius where the wind becomes supersonic
\begin{equation}
    r_c = \frac{m_pGM_\star}{4k_bT},
\end{equation}
where $G$ is the gravitational constant, $M_\star$ is the mass of the host star, $k_b$ is the Boltzmann constant, and $T$ is the temperature of the wind, equal to the coronal temperature of the host star.

And, $c_s$ is the speed of sound, which is, assuming an ideal wind of non-interacting particles, given by
\begin{equation}
    c_s = \sqrt{\pdv{P}{\rho}} = \sqrt{\frac{k_bT}{m_p}}.
\end{equation}

To resolve the degeneracy emerging from two unknowns and one equation, Equation~\ref{ParkerWind}, we make use of the expression for stellar wind speed at 1 AU quantified by the age of the host star $t$, presented in \cite{Newkirk1980},
\begin{equation}
    v(1 \, \mt{AU}, t) = v_0\left(1 + \frac{t_\star}{\tau}\right)^{-0.43}, 
    \label{Newkirk}
\end{equation}
with the constants, $v_0 = 3971\,\mt{km}\,\mt{s}^{-1}$, and $\tau=2.56 \times 10^7\, \mt{yr}$ determined by present day solar conditions.

\begin{figure}
    \centering
    \includegraphics[width=0.475\textwidth]{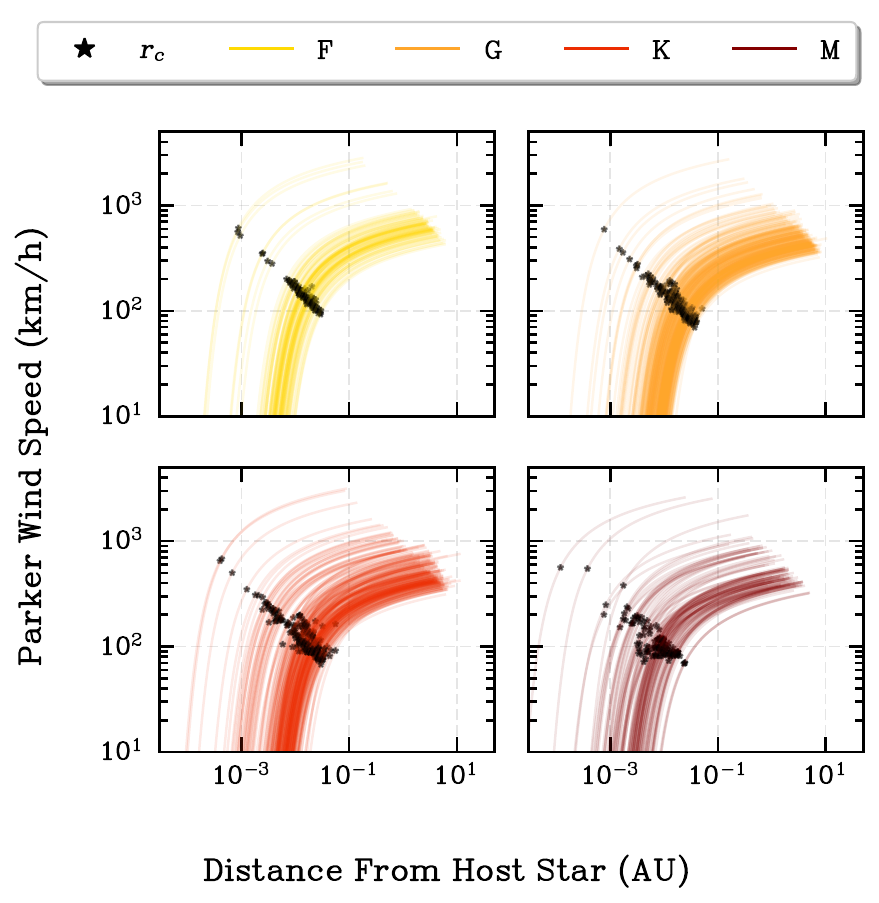}
    \caption{Isothermal wind profiles of a sample of $\sim$800 stellar systems considered in our analysis. Spectral types of the stars are indicated with line colors, and the asterisks denote the corresponding critical radii. Each profile is drawn from 0.1\textendash 200 $r_c$.}
    \label{fig: wind-profiles}
\end{figure}

Numerically solving Equation~\ref{ParkerWind} at a distance $r = 1 \, \mt{AU}$ with the coronal temperature as the free parameter to obtain the same speed resulting from Equation~\ref{Newkirk}, we infer $T$. Then, we can solve Equation~\ref{ParkerWind} with the distance as the free parameter. The resulting wind profiles for a sample of stellar systems (later to be outlined) are shown in Figure~\ref{fig: wind-profiles}.
By choosing suitable initial guesses, the numerical solutions are ensured to result in the outflow solution of the Parker wind model. This is reflected in the monotonic increase of the wind speeds with distance from the star.

The kinetic interaction between the planetary magnetosphere and the stellar wind is determined by the speed at which the planet moves through the wind. This speed, however, is not equal to the stellar wind speed since the planet is in motion around its host star. Thus, we define the effective speed of the planet,
\begin{equation}
    v_\mt{eff} = \sqrt{v^2 + v_k^2},
\end{equation}
where $v$ is the speed of the wind, which is, to a good approximation, orthogonal to the planet's orbit, and $v_k$ is the Keplerian speed of the planet, given as
\begin{equation}
    v_k = \sqrt{\frac{GM_\star}{r}},
\end{equation}
where $r$ is the distance of the planet from its host star, taken as the semi-major axis of the orbit.

Another relevant property of the stellar wind is the particle number density, which is found by the mass conservation of the star-wind system as
\begin{equation}
    n(r) = \frac{\dot{M}_\star}{4\pi r^2 m_pv(r)},
\end{equation}
with $\dot{M}_\star$ standing for the mass loss rate, which is calculated using its relation to stellar activity. Empirically, \cite{Wood2002} had determined that for solar-like GK dwarfs, the mass loss rate correlated with X-ray flux activity as $\dot{M}_\star \propto F_X^{1.15 \pm 0.20}$, and suggested that the two stars in their sample did not fit this relation because of their high activity and being less solar-like spectrum-wise. Later, with new measurements \cite{Wood2005} refined the relation as $\dot{M}_\star \propto F_X^{1.34 \pm 0.18}$, this time with an outlying GK type binary. It was concluded that this empirical relation did not hold for \textit{any} type of high-activity star, with the upper limit being $F_X \approx 8 \times 10^5 \, \mt{erg} \, \mt{cm}^{-2} \, \mt{s}^{-1}$.

Due to the lack of generality of this relation, we take an alternative route and consider the fully simulated mass loss-activity relation presented in \cite{Alvarado-Gomez2016II}. \cite{Alvarado-Gomez2016II, Alvarado-Gomez2016I} Combine their simulated mass loss-radial magnetic flux relation with the X-ray luminosity-magnetic flux relation to find
\begin{equation}
    \dot{M}_\star \propto F_X^{0.79^{+0.19}_{-0.15}},
\end{equation}
which is flatter than the empirical relations of \cite{Wood2002, Wood2005}, and it can account for young and active main-sequence stars while still explaining the discrepancy with the empirical relations \citep{Alvarado-Gomez2016II}. We take this relation and combine it with the X‐ray flux‐age relation $F_X \propto t_\star^{-1.74 \pm 0.34}$ \citep{Ayres1997} to relate the mass loss rate to stellar age as
\begin{equation}
    \dot{M}_\star \propto t_\star^{-1.37}.
    \label{eq: massloss}
\end{equation}
This relation was scaled with the Sun using the solar mass loss rate of $10^{-14}\, M_\odot /\mt{yr}$ and age of 4.603 Gyr.

\begin{figure}
    \centering
    \includegraphics{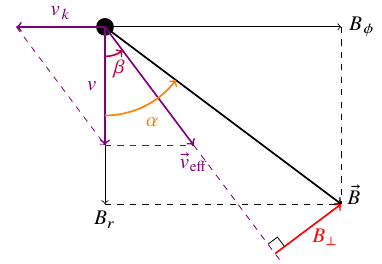}
    \caption{2D Sketch of the geometry of the interplanetary magnetic field and the related velocities, where the black dot represents the planet around its host star, which is assumed to reside in the opposite direction of $v$, the stellar wind flow. Effective wind speed incident on the planet is found from $\vec{v}_\mt{eff} = \vec{v} - \vec{v}_k$, orthogonal to which we calculate $B_\perp$.}
    \label{fig: geometry}
\end{figure}

For the IMF, in parallel with \cite{Lynch2018}, we assume a Parker spiral on the orbital plane where the components of the IMF are given by
\begin{equation}
    B_r = B_0 \left(\frac{r_\star}{r}\right)^2,
\end{equation}
and
\begin{equation}
    B_\phi = B_r \frac{\Omega_\star r}{v_\mt{eff}}.
\end{equation}
Here, $B_0$ and $r_\star$ are the surface magnetic field strength and the radius of the host star, respectively. And, $\Omega_\star$ is the angular velocity of the star with $\Omega_\star = 2\pi / P_\star$, with $P_\star$ being the rotational period of the star. Then, from the geometry sketched in Figure~\ref{fig: geometry}, the component of the IMF perpendicular to the stellar wind flow can be obtained as
\begin{equation}
    B_\perp = \sqrt{B_r^2 + B_\phi^2} \,\,\abs{\sin{(\alpha - \beta)}},
\end{equation}
with the angles
\begin{equation}
    \alpha = \arctan{\left(\frac{B_\phi}{B_r}\right)},
\end{equation}
and
\begin{equation}
    \beta = \arctan{\left(\frac{v_k}{v}\right)}.
\end{equation}

To obtain the magnitude of the perpendicular IMF strength, one must know the surface magnetic field strength of the host star, $B_0$, which can be inferred from the reconstructed stellar magnetic field via the Zeeman-Doppler imaging \citep[ZDI; ][]{Semel1989}. In their work, utilizing ZDI data, \cite{Vidotto2014} found that the unsigned large-scale magnetic field of stars correlates with stellar age via the relation
\begin{equation}
    \langle|B_V|\rangle \propto t_\star^{-0.655 \pm 0.045}.
    \label{eq: starB0}
\end{equation}
Encouraged by the observation of an agreeing trend \added{in young solar-type stars} found in a similar analysis in \cite{Folsom2016} on a narrower mass range, we scale Equation~\ref{eq: starB0} with the solar parameters $t_\odot = 4.6\,\mt{Gyr}$ and $\langle|B_V|\rangle = 1.89\,\mt{G}$ to obtain an estimate for $B_0$. \added{An important caveat associated with this approach is the different mechanisms responsible for the large-scale magnetic field strengths of fully-convective ($M\lesssim0.35\,M_\odot$) M dwarfs and radiative stars. The lack of a tachocline in such M dwarfs, which boosts magnetic fields of solar-like stars, causes their magnetic structure to be distinct \citep{Morin2008a}. In the analysis of \citet{Vidotto2014}, it can be seen that the M dwarf magnetic fields are usually underestimated by the trend in Equation~\ref{eq: starB0}. Indeed, M dwarfs are known to have complex magnetic fields with strengths reaching the kG regime \citep[e.g.,][]{Morin2008b}. Therefore, the robustness of our predictions for planets around such stars remains limited.}

\begin{figure}
    \centering
    \includegraphics[width=0.475\textwidth]{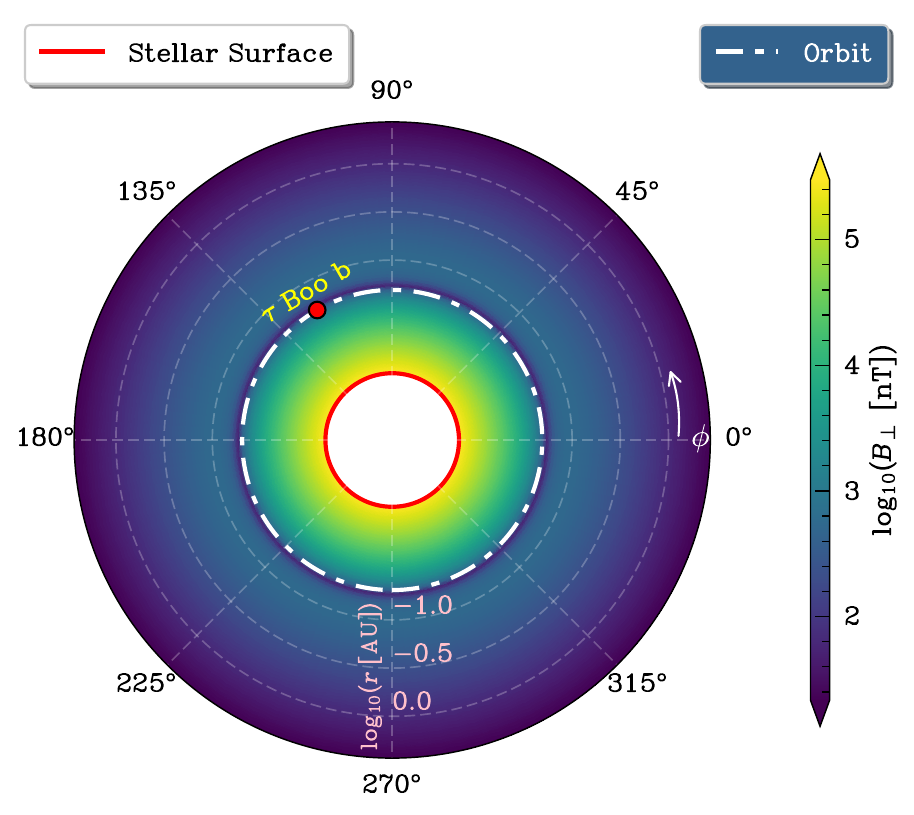}
    \caption{The perpendicular component of the IMF of the Parker spiral model for the exoplanet \mbox{tau Boo b}.}
    \label{fig: parker_spiral}
\end{figure}

The radial behavior of the perpendicular component of the IMF is exemplified for the exoplanet \mbox{tau Boo b} in Figure~\ref{fig: parker_spiral}.
It is noteworthy that right outside the orbit of the planet, $B_\perp$ has a local minimum, where it takes a value much lower than its surroundings. This physically corresponds to the condition that the IMF and the effective wind speed be nearly parallel (see Figure~\ref{fig: geometry}). Given the behavior of the wind speed (shown in Figure~\ref{fig: wind-profiles}) and its impact on $B_\phi$, such an alignment is possible. Further, depending on the rotational period of the star, this alignment could take place on either side of the planet's orbit. The existence of such a sudden drop in $B_\perp$ in the Parker model contributes to the volatility of our predictions.

\subsection{Planetary Properties}
\label{subsection: P. Properties}
The magnetospheric characteristics of exoplanets are eventually sought to be inferred from their auroral radio emissions. However, these characteristics, namely the magnetic field strength and the magnetopause standoff distance, strongly influence the nature of the CMI‐driven emissions, as seen in Section~\ref{subsection: CMI-stellar}. Therefore, an \textit{a priori} model and discussion of such planetary properties is necessary.

If the magnetic moment of the exoplanet is known, assuming the exoplanet is a dipolar magnetic structure, its magnetic field strength at its surface can be calculated as
\begin{equation}
    B_\mt{surf} = \mu \left(\frac{R_J}{R}\right)^3 \,\, [B_\mt{surf, J}],
    \label{Bs}
\end{equation}
where $\mu$ is the magnetic moment and $R$ is the radius of the planet. $B_\mt{surf, J}$ is the magnetic field strength at the surface of Jupiter and is taken to be 14 G.

Several attempts have been made to construct scaling laws based on dynamo parameters of planetary bodies \citep{Busse1976, Stevenson1983, Mizutani1992, Sano1993}. We take the relation presented in \cite{Mizutani1992},
\begin{equation}
    \mu \propto \rho_c^{1/2}\omega_p^{1/2}r_c^3\sigma_c^{-1/2},
    \label{eq: moment}
\end{equation}
where $\rho_c$, $r_c$, and $\sigma_c$ are the average density, radius, and conductivity of the convective core, respectively, and $\omega_p$ is the planetary rotation rate.

Along the lines of \cite{Ashtari2022}, for the radius of the convective core, we use two different scaling relations applying to different subsets of exoplanets. For exoplanets with masses $M < 0.4\, M_J$ and average density $\rho > 1.6 \, \mt{g}/\mt{cm}^3$, which are likely not gas giants, we use the power-law relation by \cite{Curtis1986}, which was fit with the planets Mercury, Earth, Jupiter, and Saturn,
\begin{equation}
    r_c \propto M^{0.44}.
    \label{rc1}
\end{equation}
As they suggest, this relation might be viewed as a state equation for the internal dynamics of the planet. In younger, non-equilibrium systems where the radius of the planet is not well determined only by its mass but also its age, we use the relation scaled for solar system gas giants in \cite{Griessmeier2004} as
\begin{equation}
    r_c \propto M^{0.75} R^{-0.96}.
    \label{rc2}
\end{equation}
This latter case is typical for Hot Jupiters (HJs), as they tend to have very low densities and are found mostly around younger stars \citep{Chen2023}. We scale Equations~\ref{rc1} and \ref{rc2} with the Jovian convective dynamo region radius of $0.830 \, R_J$ reported in \cite{Sharan2022}, found by modeling the Jovian magnetic field based on Juno data.

Due to the lack of any analytical method to predict them, the densities of the convective cores were assumed to be the same as the mean densities of the corresponding bodies for every exoplanet. Similarly, the conductivities of the convective cores for every exoplanet were assigned that of Jupiter.

Since the pioneering work of \cite{Snellen2014}, where they measured the rotation rate of a gas giant extrasolar planet for the first time using spectral Doppler analysis of the planet's transit data, numerous similar measurements took place for other exoplanets \citep{Brogi2016, Schwarz2016, Xuan2020, Wang2021}. However, since this method relies on high-resolution transit observations that are not widely applicable, the rotation rates of exoplanets remain elusive. Therefore, they were estimated in this work in the following manner. Exoplanets with $a < 0.1 \, \mt{AU}$ were assumed to be tidally synchronized. Thus, their orbital periods were assigned as their rotation periods. A statistical approach was implemented for the remaining exoplanets. Their rotational angular momenta were randomly sampled from a probability density function obtained from a kernel density estimation on the momenta of solar system planets as part of the Monte Carlo error propagation. Their rotation rates were then recovered from their respective momenta. All extrasolar planets were assumed to be spherical bodies of homogeneous mass distribution, so the angular momentum relation is
\begin{equation}
    L = I\omega_p = \frac{2}{5}MR^2 \omega_p.
\end{equation}

Having determined a value for the planetary magnetic field and identified the required stellar wind properties, we fix the magnetopause distance of the exoplanet's magnetosphere by considering the force balance at the magnetosphere boundary. In a similar manner to \cite{Lynch2018}, we consider the magnetic pressures caused by the planet and the IMF, along with the stellar wind thermal and ram pressure,
\begin{equation}
    \frac{B(r)^2}{2\mu_0}= \frac{B_\star^2}{2\mu_0} + 2 n k_B T + m_p n v_\mt{eff}^2.
\label{pressure eq.}
\end{equation}
Here, $B(r)$ is the planetary magnetic field at the location of the magnetopause, $B_\star$ is the IMF strength at the exoplanet location given simply as $B_\star = \sqrt{B_r^2 + B_\phi^2}$, $\mu_0$ is the permeability of free space, and other symbols have their usual meanings. Assuming a magnetic dipole, planetary magnetic field strength scales with the inverse cube of the distance from its origin. Hence, one can rearrange Equation~\ref{pressure eq.} to obtain the magnetopause distance as
\begin{equation}
    \frac{R_\mt{mp}}{R} = \left[ \frac{(2.44 \times B_\mt{surf})^2}{2\mu_0(\frac{B_\star^2}{2\mu_0} + m_p n v_\mt{eff}^2 + 2 n k_B T)}\right]^{1/6},
\end{equation}
where $R$ is the planetary radius. The factor 2.44 arises in the Mead-Beard magnetosphere model due to magnetospheric current fields \citep{Mead1964}. 

\section{Methodology \& the Exoplanet Sample}
\label{section: Methodology}
\begin{figure*}[htbp]
    \centering
    \includegraphics[width= \textwidth]{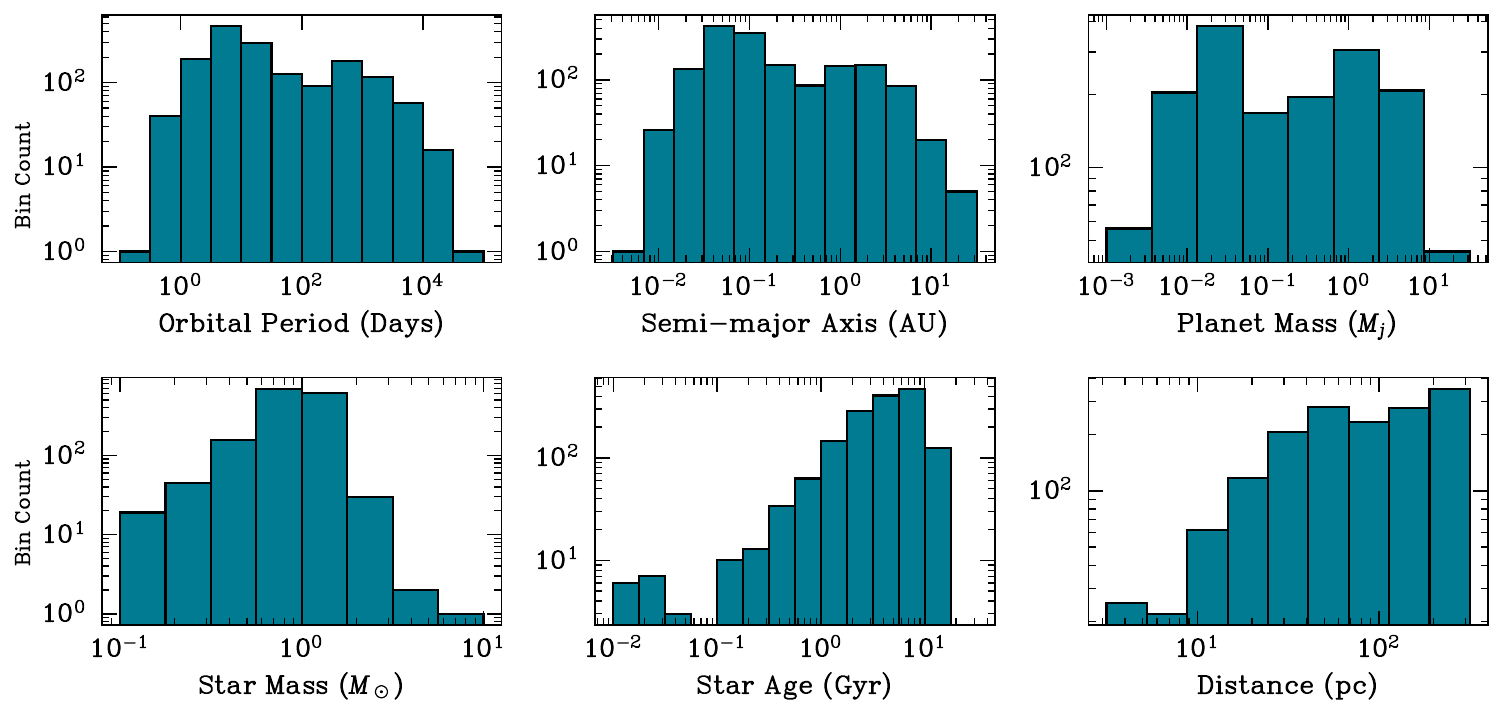}
    \caption{Distributions of \added{some of }the parameters of the exoplanets and their host stars in our sample.}
    \label{fig: initial}
\end{figure*}

The sample of confirmed exoplanets and planetary system parameters were taken from the NASA Exoplanet Archive \added{\citep{NAE_PS_Composite}}. Out of the \added{5885} confirmed exoplanets available in the Planetary Systems Composite Data table as of \added{April 30, 2025}, those that lacked either one of the orbital period, semi-major axis, planetary radius, mass, and density; host star's, stellar radius, mass, and age; and system distance were left out,\added{ and a maximum distance cut was placed at 300 pc,} bringing the number of exoplanets to \added{1637}. Then, we removed \added{40} targets with masses $M > 13\, M_J$. Focusing on main-sequence and pre-main-sequence dwarf stars, we also restricted the spectral type of hosts to include only F, G, K, and M-type stars. \added{To make this assertion, we placed an upper limit on the effective surface temperature of the host at 7000 K, and manually removed HAT-P-57, which was still classified as an A-type star.} Overall, this  removed \added{17} planets. With its extreme orbital distance of 7506 AU, COCONUTS-2b is also removed from our sample since its unusual circumstance renders it incompatible with our simple stellar wind model. After all these considerations, we end up with \added{1579} suitable exoplanets. The distributions of a subset of the parameters of selected exoplanets are shown in Figure~\ref{fig: initial}. 

\begin{figure}[htbp]
    \centering
    \includegraphics[width = 0.475\textwidth]{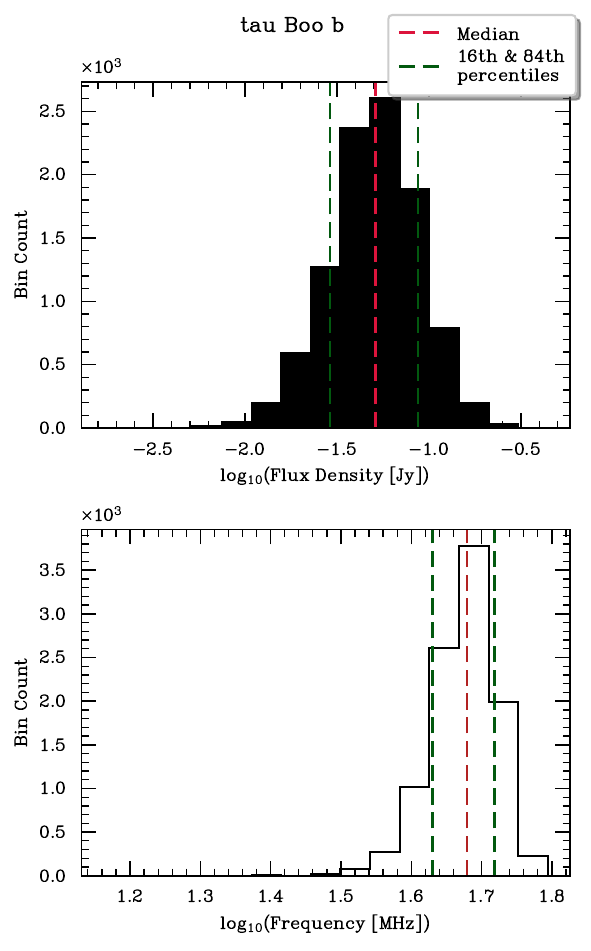}
    \caption{Distribution of the frequency and the flux density of the expected radio emissions from \taub{} as an outcome of the error propagation method. The sixteenth and the eighty-fourth percentile lines correspond to the reported error bars of our predictions.}
    \label{fig: taub}
\end{figure}

Using the model outlined in Section~\ref{section: Model}, we determine the expected radio emission frequencies and flux densities of the planets in the sample. In doing so, a numerical Monte Carlo method is implemented to account for uncertainties. The uncertainties on the planetary system parameters are taken from the exoplanet catalog if available. For those without reported uncertainties, we assume a conservative uncertainty of 20\% of the catalog value. For any empirical/simulated power law relation in the model, we also consider the uncertainty of the exponents. Then, we implement a Monte Carlo error propagation method, sampling over the assumed normal distributions of the planetary system parameters and the power law exponents. As a result, we obtain, for each exoplanet, a posterior probability distribution for its emission frequency and flux density. An example of these distributions for the case of \taub{} is shown in Figure~\ref{fig: taub}.

We find that $\sim \added{470} \, (\added{30}\%)$ of the exoplanetary CMI emission fail to escape from the local plasma of their environment and thus cannot reach the Earth. The results for the remainder of our sample are analyzed in Section~\ref{subsection: outcomes}.

To assess the detectability of the predicted radio emissions, we consider four observing facilities. The sensitivity limits of the Low-Band Antenna (LBA) and High-Band Antenna (HBA) arrays of the Low-Frequency Array (LOFAR) are taken from \cite{LOFAR}. In contrast, those of the Murchison Wide Field Array (MWA) and the upgraded Giant Metrewave Radio Telescope (uGMRT) are obtained from \cite{MWA} and \cite{uGMRT}, respectively. Finally, the thermal noise levels of the New Extension in Nançay Upgrading LOFAR (NenuFAR) were obtained through calculations of its system equivalent flux density (SEFD) throughout its wavelength range at the zenith, with critical data and algorithms being the courtesy of Dr. Philippe Zarka through private communication. The reported RMS noise levels for each band in every observatory are scaled to an optimistic integration time of 8 hours. However, the integration times are practically limited by the actual positioning of sources in the sky with respect to different telescopes. We evaluate the elevation of a given source on the celestial sphere,
\begin{equation}
    \sin{\phi} = \sin{\lambda}\sin{\delta} + \cos{\lambda}\cos{\delta}\cos{(\alpha_{LST} - \alpha)},
\end{equation}
whose right ascension ($\alpha$) and declination ($\delta$) are known at latitude $\lambda$ on the surface of the Earth, where $\alpha_{LST}$ is the Local Sidereal Time (LST). Numerically solving for $\phi$ for a 24-hour $\alpha_{LST}$ series, we obtain the fraction of a day in which the source is above an arbitrary level of elevation. This is then used to evaluate the possible observation times of targets using different telescopes.

\section{Results}
\label{section: Results}

\subsection{Model Outcomes}
\label{subsection: outcomes}
\begin{figure*}[htbp]
    \centering
    \includegraphics[width = \textwidth]{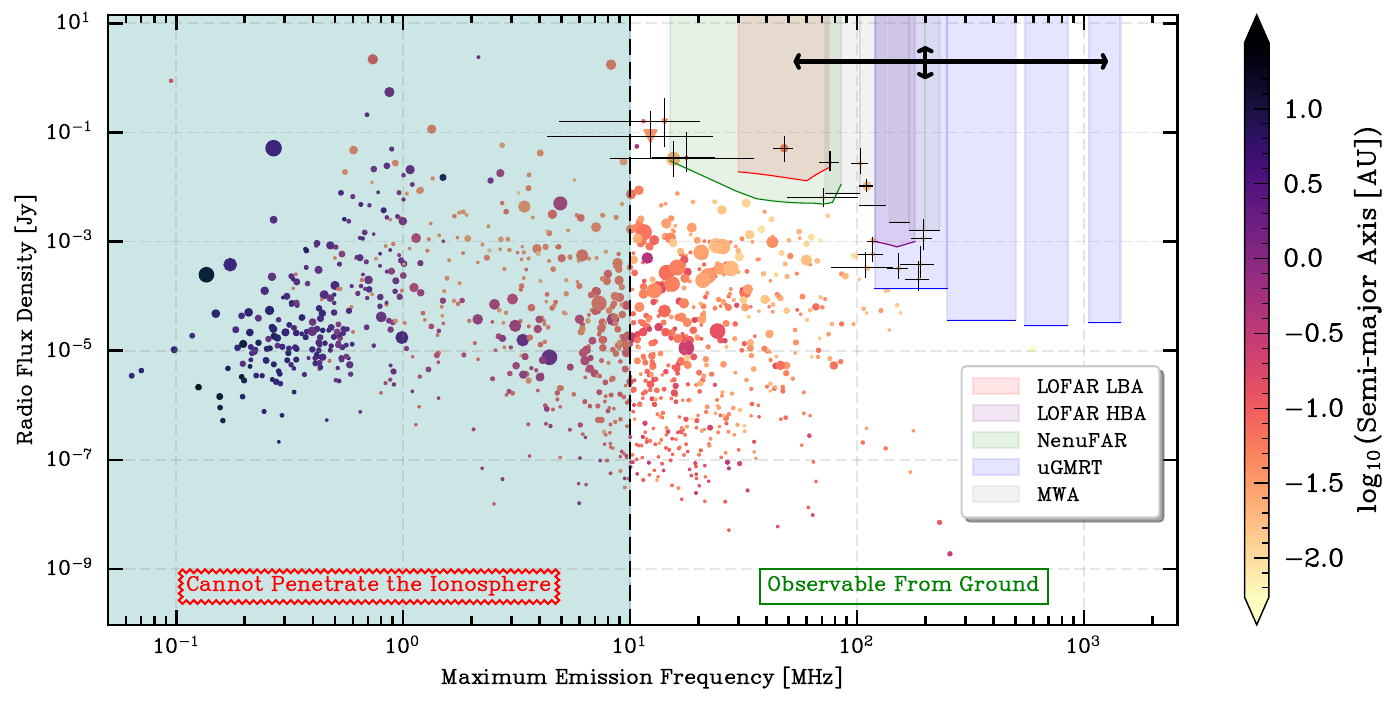}
    \caption{The burst CMI‐driven magnetospheric emission frequency and Earth-distance radio flux density of a sample of $\added{1111}$ Exoplanets. The pair of thick horizontal and vertical error bars at the top right show the average uncertainty. Dot sizes indicate the distance from Earth to the exoplanets with bigger dots indicating closer systems. Earth's ionospheric cutoff is shown at 10 MHz. 5$\sigma$ sensitivities of the observatories LOFAR, NenuFAR, MWA, and uGMRT are indicated, all scaled to a common integration time of 8 hours.}
    \label{fig: scatter}
\end{figure*}

\begin{figure*}[htbp]
    \centering
    \includegraphics[width=\textwidth]{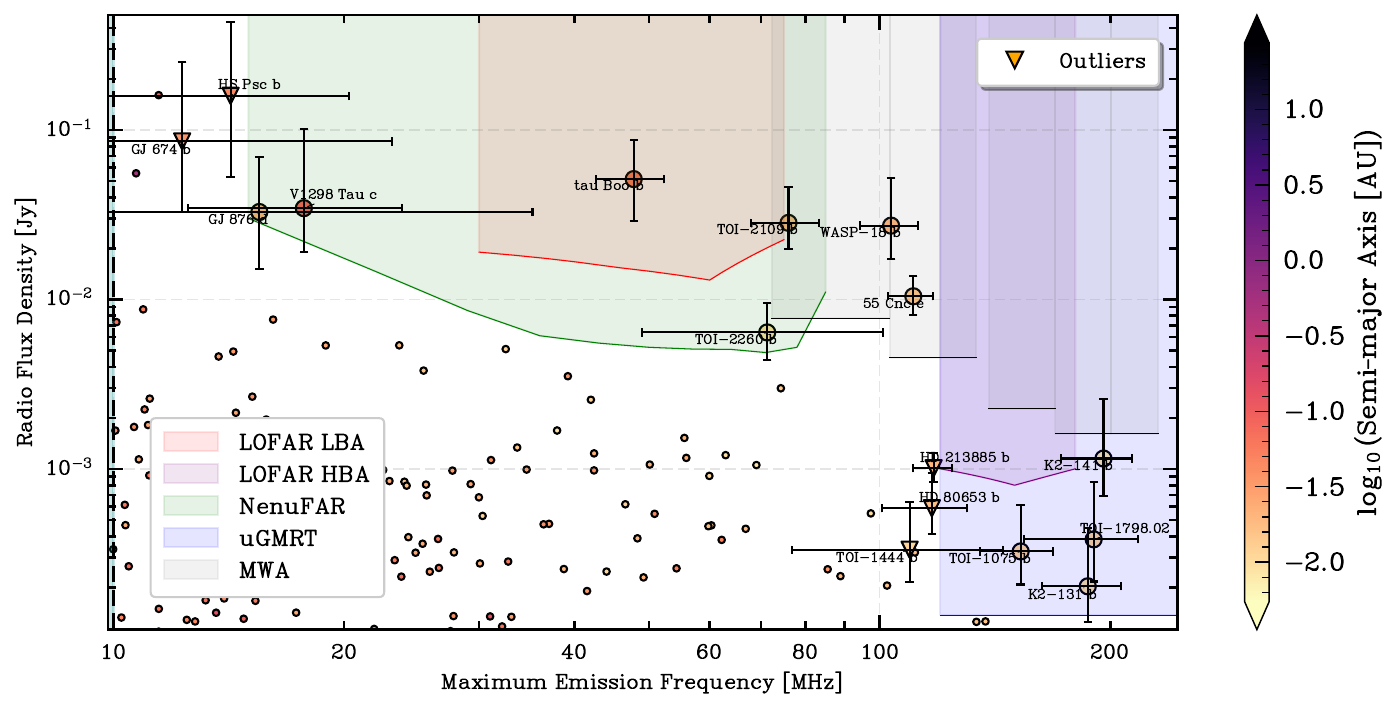}
    \caption{Section of Figure~\ref{fig: scatter} that shows the error bars and labels for the top candidates with the highest predicted observational relevance. Dot size differentiation is given up to enhance the visual. Candidates with adequately high flux densities whose median posterior frequencies lie out of the observation bands despite being very near them are dubbed "outliers" and indicated with the upside-down triangles.}
    \label{fig: zoom}
\end{figure*}

Using the outlined exoplanet sample and the described model, the maximum radio emission frequency and flux density were obtained for $\added{1111}$ exoplanets, whose radio emission is expected to penetrate through the local wind plasma in their environment. Figure~\ref{fig: scatter} presents the predicted Earth-distance radio flux density and the frequency of the radio emission of the sample. A closer look into the exoplanets whose emissions are at the suitable frequencies and have flux densities above the respective sensitivity limits of the considered observing facilities is provided in Figure~\ref{fig: zoom}.
The uncertainties for the maximum emission frequency and flux density values were obtained from each system's respective posterior distributions (see Section~\ref{section: Methodology}). The individual uncertainties for the top candidates are indicated in Figures~\ref{fig: scatter} and~\ref{fig: zoom}. For the remaining systems, Figure~\ref{fig: scatter} provides a single uncertainty averaged over all exoplanets. We find that, on average, the predictions for the frequencies have uncertainties of a factor of 4 toward lower values and of a factor of 7 toward higher values. On the other hand, the flux densities have an uncertainty of about a factor of 2.5 toward both brighter and dimmer values. These are indicated by the dark cross on the top right corner of Figure~\ref{fig: scatter}. Hence, we note that the precision of our frequency estimates is lower than those estimated by \cite{Lynch2018}, but our flux densities have lower uncertainties.

Our predictions span ten orders of magnitude in radio flux densities and three orders of magnitude in peak frequency. In the set of emitting planets whose radiation can escape their local plasma, $\sim$ \added{530} (\added{48}\%) are predicted to have CMI-driven radio emissions with frequencies above 10 MHz and can propagate through the Earth's ionosphere. Another observation is that besides $\sim \added{90}$ exceptions, these bodies are those assumed to be tidally synchronized to their host star and are thus faster rotators than those whose spin angular momenta were sampled from the solar system bodies. Of these, \added{16} have high enough radio flux \added{densities} at Earth at the correct frequencies that render them observable by the facilities considered. Correct frequencies refer here to the distribution of the maximum emission frequency of the exoplanet having fallen at least 30\% into one of the observation bands of the facilities. 

For those deemed observable, the provided error bars signify the sixteenth and eighty-fourth percentiles of the respective posterior distributions, which quantify the uncertainty in the predictions as an output of the Monte Carlo error-propagation method. This percentiles method allows for asymmetric error bars, and the percentile limits are those obtained from the standard deviation of the normal distribution. The outcome is that many of these reveal expected detection significances of a few $\sigma$s. It should be noted, however, that although for the unreported uncertainties in the planetary system parameters, a conservative detection significance of $5\sigma$ was assumed, no similar thing was done for unreported uncertainties in the exponents of the empirical power law relations. This is one of the many reasons, along with the strong assumptions of the model, why the predictions reported here should be taken as a guide for future observation programs and not strong assertions.

As illustrated in Figures~\ref{fig: scatter} and~\ref{fig: zoom}, there are \added{five} exoplanets whose data points are not contained within the considered observation bands, namely \mbox{\added{HS Psc b}}, \mbox{\added{GJ 674 b}}, \mbox{HD 80653 b}\added{, \mbox{TOI-1444 b}} and \mbox{HD 213885 b}. These are those whose median values in the distributions of their flux densities and maximum emission frequencies do not fall into any bands. However, since their frequency distributions lie within at least 30\% of any of the boxes and are predicted to have sufficiently high flux densities for the corresponding box, we have chosen to present them as possible targets for observation campaigns. \added{First two} of these "outliers" fall into \mbox{\added{NenuFAR}}, \added{while the rest fall into} \mbox{uGMRT}.

\added{All} of the candidates are "hot" planets that have orbits with small semi-major axes. \added{Some} of the brighte\added{r ones}, \added{\mbox{HS Psc b},} \taub{}, \added{\mbox{V1298 Tau c,}} \wb{}, and \mbox{TOI-2109 b} are \added{gas giants}, while most of the\added{ remaining} are super-Earth-type planets. Such a domination of super-Earths is extraordinary in the literature—partly reflecting how better we have gotten at detecting them. Among the candidates, the systems \mbox{55 Cnc} and \mbox{tau Boo} are old and popular planetary radio emission candidates.

\subsection{Observability Status of the Candidates}
\label{subsection: obs}
\begin{figure*}[htbp]
    \includegraphics[width = 0.95\textwidth]{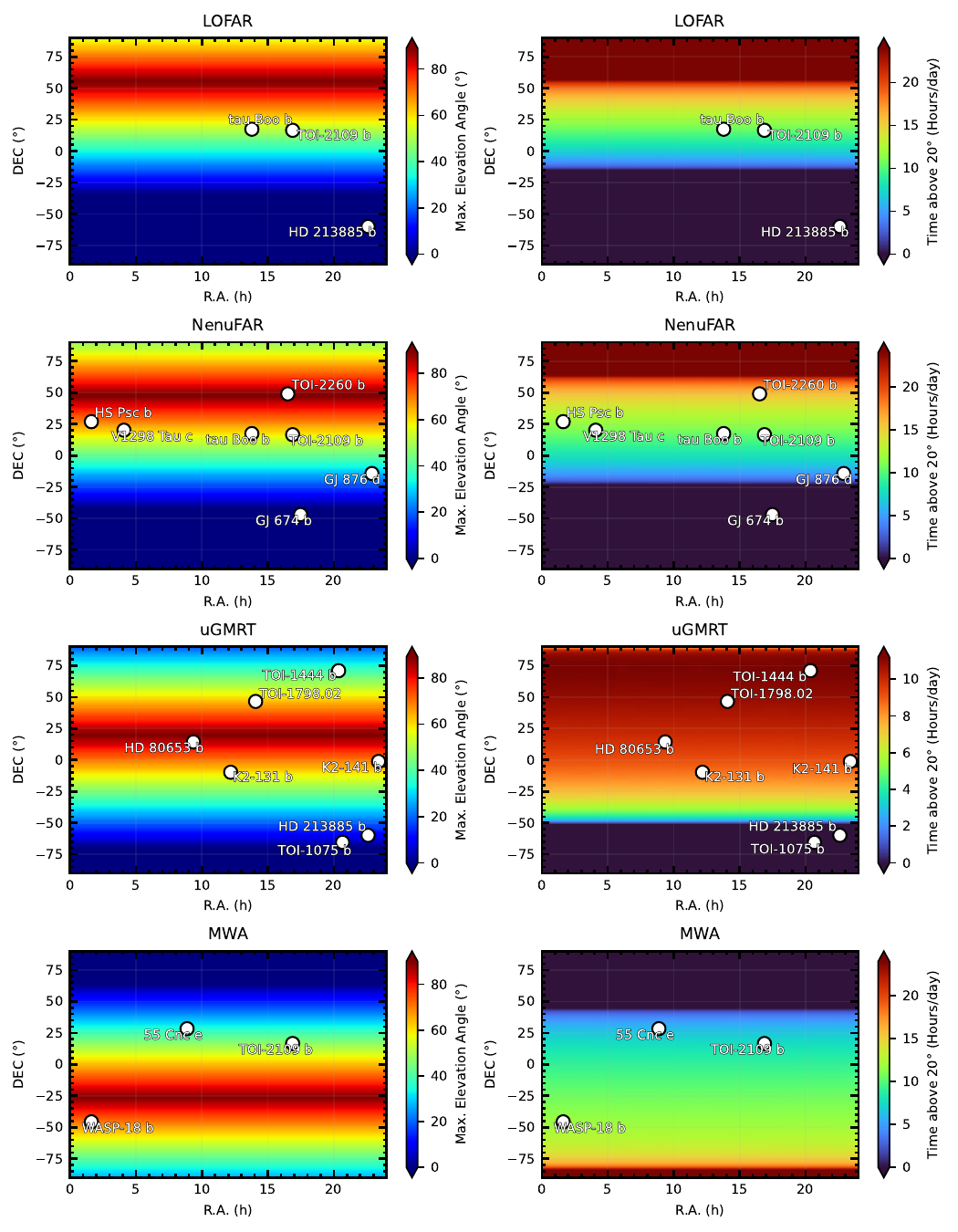}
    \caption{Sky coverage maps of the considered observatories LOFAR, NenuFAR, MWA, and uGMRT. Left: Plots for the maximum elevation of each point in the observer's sky, Right: Time spent above 20$\degree$ altitude at the observer's sky. In all panels, the candidate exoplanets are marked with white dots.}
    \label{fig: maps}
\end{figure*}
To further assess the observability of the candidate exoplanets, one must consider the sky coverage of the observing facilities. For this purpose, the maximum elevation maps and the respective possible observation duration maps are provided in Figure~\ref{fig: maps}. 
Optimizing the antenna radiator diagram for \mbox{NenuFAR}, \cite{Zarka2015b} found significant extinction in the electronic beam-forming for sources below 20$\degree$ elevation on the sky. This is why we take this as the minimum elevation for beam-formed observations.

It is apparent that its \added{southern}most target candidate \mbox{HD 213885 b} never rises above 20 degrees at LOFAR. \added{Further}, the uGMRT is blind to its southern candidates \mbox{HD 213885 b} and \mbox{TOI-1075 b}\added{, and NenuFAR to \mbox{GJ 674 b}}. Besides these limitations, the facilities can observe the sources for several hours. To more clearly present observability results and provide a closer look into the exoplanets found to be in the observable range by our model, we present the parameters and the model results of the candidates in Table~\ref{tab: obs}. Also included in there are lists of telescopes with which the candidates are deemed visible, ordered by decreasing possible integration time. The times the candidates spend above 20$\degree$ elevation at the telescopes are calculated, and the longest time is also reported in the last column of Table~\ref{tab: obs}.

\begin{table*}[htbp]
    \centering
    \caption{Table of the exoplanets found candidates for having detectable radio emission driven by CMI. The celestial coordinates, five basic parameters, the results for maximum emission frequency, and the radio flux density of the exoplanets are provided. The last two columns are reserved for observability and show the respective telescopes for the target and the time $t_{20\degree}$ the target spends above 20$\degree$ elevation at the position of the most suitable facility, respectively.}
    \begin{tabular}{p{2cm} p{1.2cm} p{1.5cm} p{1cm} p{1cm} p{1cm} p{0.5cm} p{1cm} p{1cm} p{1cm} p{1.2 cm} p{1.2 cm}} \hline \hline
        Name & RA (J2000) & DEC (J2000) & Mass ($M_J$) & Radius ($R_J$) & $a$ ($10^{-2}$ AU) & $d$ (pc) & $t_\star$ (Gyr) & $\nu_\mt{max}$ (MHz) & $\Phi_\mt{peak}$ (mJy) & Telescope* & $t_{20\degree}$ \\ \hline
        HS Psc b & 01:37:23 & +26:57:10 & 1.460 & 1.21 & 4.3 & 38 & 0.13 & 14 & 158.90 & 2 & 11h 54m \\ 
        GJ 674 b & 17:28:41 & -46:53:56 & 0.035 & 0.30 & 3.9 & 5 & 0.55 & 12 & 85.59 & 2 & 0h 0m \\ 
        tau Boo b & 13:47:15 & +17:27:26 & 5.950 & 1.14 & 4.9 & 16 & 2.00 & 48 & 51.27 & 1, 2 & 10h 38m \\ 
        V1298 Tau c & 04:05:20 & +20:09:25 & 0.240 & 0.50 & 8.3 & 108 & 0.02 & 18 & 34.54 & 2 & 10h 57m \\ 
        GJ 876 d & 22:53:18 & -14:16:00 & 0.021 & 0.22 & 2.1 & 5 & 1.00 & 16 & 32.78 & 2 & 4h 58m \\ 
        TOI-2109 b & 16:52:45 & +16:34:48 & 5.020 & 1.35 & 1.8 & 262 & 1.77 & 76 & 28.28 & 1, 2, 4 & 10h 28m \\ 
        WASP-18 b & 01:37:25 & -45:40:40 & 10.200 & 1.24 & 2.0 & 123 & 1.57 & 103 & 27.24 & 4 & 11h 46m \\ 
        55 Cnc e & 08:52:35 & +28:19:47 & 0.025 & 0.17 & 1.5 & 13 & 10.20 & 111 & 10.43 & 4 & 6h 0m \\ 
        TOI-2260 b & 16:30:40 & +49:02:48 & 0.010 & 0.14 & 1.0 & 101 & 0.32 & 71 & 6.38 & 2 & 15h 50m \\ 
        K2-141 b & 23:23:40 & -01:11:21 & 0.016 & 0.13 & 0.7 & 62 & 6.30 & 196 & 1.15 & 3 & 9h 7m \\ 
        HD 213885 b & 22:35:56 & -59:51:53 & 0.028 & 0.16 & 2.0 & 48 & 3.80 & 118 & 1.01 & 1, 3 & 0h 0m \\ 
        HD 80653 b & 09:21:21 & +14:22:04 & 0.018 & 0.14 & 1.7 & 110 & 2.67 & 117 & 0.58 & 3 & 9h 48m \\ 
        TOI-1798.02 & 14:04:23 & +46:31:09 & 0.018 & 0.13 & 1.1 & 113 & 2.60 & 190 & 0.38 & 3 & 10h 45m \\ 
        TOI-1444 b & 20:21:54 & +70:56:37 & 0.011 & 0.13 & 1.2 & 125 & 3.80 & 109 & 0.33 & 3 & 11h 12m \\ 
        TOI-1075 b & 20:39:53 & -65:26:59 & 0.031 & 0.16 & 1.2 & 61 & 6.00 & 153 & 0.33 & 3 & 0h 0m \\ 
        K2-131 b & 12:11:00 & -09:45:55 & 0.025 & 0.15 & 0.9 & 153 & 5.30 & 187 & 0.20 & 3 & 8h 37m \\   \hline
        \multicolumn{12}{l}{* The telescopes are numbered as 1: LOFAR, 2: NenuFAR, 3: uGMRT, 4: MWA}
    \end{tabular}
    \label{tab: obs}
\end{table*}

It is essential to keep in mind that the sensitivity limits of the facilities were determined with the assumption of an integration time of 8 hours. This integration time is not feasible for certain candidates and the respective telescopes, as will be discussed in Section~\ref{subsection: Opportune Targets}.

Additionally, although scattered and absorbed by Earth's ionosphere, radio waves with frequencies below 10 MHz from exoplanets can have high flux densities at the location of Earth. We report our model's brightest five \added{such} candidates with their celestial coordinates and emission characteristics in Table~\ref{tab: space-obs2}.

\begin{table}[htbp]
\centering
\caption{Celestial coordinates and emission characteristics of the brightest candidates in the frequency range 0.1–10~MHz, for future space-based observations. It should be noted that the emission characteristics contain large uncertainties.}
\begin{tabular}{p{1.7cm} p{1.2cm} p{1.3cm} p{1cm} p{1.2cm}} \hline \hline
        Name & RA (J2000) & DEC (J2000) & $\nu_\mt{max}$ (MHz) & $\Phi_\mt{peak}$ (Jy) \\ \hline
        HD 36384 b & 05:39:44 & +75:02:38 & 2.15 & 2.40 \\ 
        AU Mic c & 20:45:10 & -31:20:33 & 0.73 & 2.19 \\ 
        AU Mic b & 20:45:10 & -31:20:33 & 8.24 & 1.74 \\ 
        HD 62509 b & 07:45:19 & +28:01:34 & 0.87 & 0.55 \\ 
        bet Cnc b & 08:16:31 & +09:11:07 & 0.69 & 0.21 \\    \hline
    \end{tabular}
    \label{tab: space-obs2}
\end{table}

An extended version of Table~\ref{tab: space-obs2} containing the emission characteristics with their corresponding uncertainties of all of the exoplanets in our model are available in machine-readable format.

\section{Discussion}
\label{section: Discussion}
The overall range of the outcomes of our model is in good agreement with the literature, including the latest works by \cite{Lynch2018}, \cite{Ashtari2022}, and \cite{Mauduit2023a}. In contrast to the former, we do not predict as many exoplanets with maximum emission frequencies above 100 MHz. This is due to the difference between the scaling laws employed therein and in this work regarding the magnetic field strength of substellar companions. In particular, while we use the scaling relation in \citep{Mizutani1992} (Equation~\ref{eq: moment}), \cite{Lynch2018} had adapted the dynamo model of \cite{Reiners2010}, which scales the magnetic field strength with the planetary mass, radius, and luminosity.

Although in better alignment overall, our results differ significantly from those of \cite{Ashtari2022} for some of the targets. For instance, the maximum emission frequency of \taub{} in our results is more than twice the value they predict, which is 19 MHz. More substantially, our result for the flux density from \mbox{AU Mic c} is about \added{five} times larger than their predictions for high activity burst emission. However, the frequency we found is smaller by \added{a factor of $\sim$35}, rendering \mbox{AU Mic c} unobservable from the ground. Despite most scaling laws being common in both works, these differences stem mainly from the different methods used to model different stellar environments. \cite{Ashtari2022} had employed the main sequence scaling laws of \cite{Johnstone2015} to account for the stellar wind parameters simply in terms of the mass and the mass loss rate of the host star. In contrast, we assumed isothermal Parker wind for each star and proceeded to calculate the wind parameters separately. This further allowed us to determine the magnetopause distance directly from the force balance at the magnetosphere boundary while they had to rescale that of Jupiter, assuming solar wind parameters. With a smaller impact, some of our scaling laws had slightly different initial conditions. For instance, we have used the recent value of $0.830\,R_J$ \citep{Sharan2022} for the radius of the convective dynamo region of Jupiter instead of the canonical $0.9\,R_J$. 

When compared to \citet{Mauduit2023a} who employed a similar formalism with this work the predictions are mainly consistent. However, significant differences for several targets are again present. The main source of this dissimilarity is the variety of scaling laws regarding the planetary magnetic moment (Equation~\ref{eq: moment}), over which \cite{Mauduit2023a} performed a geometric average (viz. those in \cite{Busse1976} and \cite{Sano1993} along with \cite{Mizutani1992}). Furthermore, their use of the dynamo model of \cite{Reiners2010} likely gave rise to higher predicted frequencies than was found in this work, similar to the case of \cite{Lynch2018}.

Finally, we have accounted for the uncertainties of inputs for each system, which may have given rise to differences with the previous findings of such predictive works. The number of Monte Carlo iterations our model can implement remains limited by the increasing computational power required. Hence, the results for a few systems with highly uncertain planetary system parameters tend to be unstable. On average of multiple runs, we have found that with $\added{10^{\added{4}}}$ Monte Carlo iterations, there are \added{13} exoplanets whose posterior median emission frequencies deviate at least by a factor of 1.5 from what one would obtain had the uncertainties not been included. Due to the impact of the emission frequency on the flux density through Equation~\ref{eq: flux}, the same exoplanets also tend to have flux densities of such deviation. As a result, there are \added{19} systems with such high deviation of flux density. 

\subsection{Opportune Targets}
\label{subsection: Opportune Targets}
Among the favorable systems for ground-based observations, considering the distances to them given in Table~\ref{tab: obs}, the top \added{four} most intensely emitting are \mbox{TOI-2109 b}\added{, \wb{}, \mbox{V1298 Tau c}, and \mbox{HS Psc b},} which are all gas giants. They are followed by the super-Earth \mbox{TOI-2260 b}, which has an emission power per unit frequency more than ten times that of the HJ that follows, \taub{}. This intense emission from a super-Earth is due to the young age of \mbox{TOI-2260}, which is $321\pm96$ million years old \citep{Giacalone2022}. Similarly, the intense emission\added{s} from \mbox{V1298 Tau c}\added{ and }\mbox{\added{HS Psc b}} \added{are} also likely due in most part to the young ages of \added{these} systems, $23\pm4\,\mt{Myr}$, \added{and,} $\added{133}_{-20}^{+15}\,\mt{Myr}$ \added{respectively, \citep{David2019, Gagne2018}}. Our model favors such young systems by assigning the host star a more violent mass loss rate (Equation~\ref{eq: massloss}) and a stronger interplanetary magnetic field (Equation~\ref{eq: starB0}). Indeed, the remaining super-Earths have emissions less intense than all of the HJs, with \mbox{\added{GJ 876 d}} sustaining the smallest power per unit frequency, more than \added{2,500} times smaller than the most luminous candidate \mbox{TOI-2109 b}. 

Except \ffc{}\added{, \mbox{K2-131 b}, \mbox{GJ 674 b}, and \mbox{GJ 876 d},} the \added{remaining seven} super-Earths among our top candidates were all discovered in the last six years, all by the transit method. \added{One} of these detections, \added{that} of \mbox{K2-141 b}, were achieved by Kepler's K2 mission. In the discoveries of the other \added{six} planets, TESS observations played a key role enabled by the full-sky coverage of TESS \citep{Guerrero2021}. 
As the number of yet-to-be-confirmed TESS exoplanet candidates exceeds several thousand, interesting super-Earths and sub-Neptunes are getting more in reach, revealing a new subset of candidates for magnetospheric radio emissions. Although predicted to have sufficient flux densities at the correct frequencies, not all such candidates are positioned in the sky so that they can be observed by the respective telescope, as discussed in Section~\ref{subsection: obs}.\added{ In such cases, it might be worthwhile to consider different observing facilities for the targets. For instance}, an alternative telescope for the uGMRT candidates that never rise above 20$\degree$ could be MWA. However, the $t_{20\degree}$ values for \mbox{HD 213885 b} and \mbox{TOI-1075 b} \added{with MWA} are 12h 48m and 13h 24m, respectively, which are not enough to boost the sensitivity of the MWA to detect emission from these sources, even if observation times were granted for $t_{20\degree}$.

Model predictions remain unconstrained without experimental data, challenging the search for radio signals from exoplanets. Therefore, predictive models should be tested with the upper limits on the radio flux densities of exoplanets so far observed. The non-detection of emissions from our top candidates can be explained in several ways. First, some observations are performed at frequencies higher than the maximum emission frequency we predict, in which case, there is no disagreement with our predictions because the observation is irrelevant. Secondly, since the CMI-driven emissions are beamed, Earth might be lying outside of the emission cone of the exoplanet during the full or some part of the observations. Another explanation could be that during the observation, the exoplanet might have an orbital phase such that its emission needs to propagate through denser plasma regions in its stellar system compared to its orbit, where the plasma absorbs the entire or some part of the emission spectrum. Although this last effect can be accounted for in targeted searches, it should be considered when dealing with non-detections from larger campaigns like all-sky surveys. We now consider the observation histories of our candidates and compare them with our findings. 

\subsection{Comparison to Observations}
\label{subsection: comparison}
Encouraged by the Voyager/PRA (Planetary Radio Astronomy) observations of the spectra of the auroral emission from five solar system planets—Earth, Jupiter, Saturn, Uranus, and Neptune presented in \cite{Zarka1992}, we consider relevant the upper limits derived from observations at frequencies up to an order of magnitude lower than our predicted maximum emission frequency. Of the \added{16} sources we find to fall in the observable region, only \added{four} systems have been observed under suitable conditions: \ffc, \taub, \wb\added{, and \mbox{GJ 876 d}}. These are now going to be discussed in detail. 

Our prediction of radio emission with a flux density of about $\added{10.4}$ mJy at $111$ MHz for \ffc{} is consistent with the $3\sigma$ upper limit of 28 mJy at $150$ MHz placed by \cite{Sirothia2014} through GMRT observations, since the limit was placed for a higher frequency than our predicted maximum. Further, the campaign of \cite{Turner2021} resulted in an upper limit of $73$ mJy, for which the observations were carried out in the range 26–73 MHz with \mbox{LOFAR} LBA. Therefore, we emphasize the need for more sensitive observations below \added{111} MHz to detect CMI-driven emissions from \ffc{}. As shown in Figure~\ref{fig: zoom}, we find such campaigns would be possible with the MWA, given an integration time of 8 hours. However, we find that \ffc{} spends only 6 hours above 20$\degree$ elevation, as seen from MWA. This would scale our calculated thermal noise of MWA up by a factor of $\sqrt{4/3} \approx 7/6$, still rendering the emission from \ffc{} visible.

\cite{Lynch2018} performed a targeted search in the southern all-sky Stokes V observation data of \cite{Lenc2018} at 200 MHz with MWA and placed upper limits on the radio emissions of 18 exoplanetary systems. Included in there was \wb, with a $3\sigma$ flux density limit of 4.1 mJy. Although the resulting upper limit is lower than our prediction of \added{27} mJy, it is still consistent with our results. We do not expect to detect magnetospheric emissions from \wb{} at frequencies above 104 MHz. \added{Similarly, they placed an upper limit of 4.5 mJy for the \mbox{GJ 876} system, and there also exists an upper limit of 17.3 mJy at 150 MHz placed by \citet{Murphy2015}. These limits, although lower than our expectation at 33 mJy, do not clash with our predictions as we predict a frequency upper-bound at 16 MHz.} We encourage new, targeted observations of \wb{} with the same telescope at lower frequencies. High integration times of observation should be possible given the extended time $t_{20\degree} =$ 11h 46m.

\taub, being an opportune target for radio observations, has also been the subject of several observation campaigns for over two decades \citep{Farrell1999}. In recent years, \cite{Hallinan2013}, following their non-detection with GMRT at 150 MHz, had placed a $3\sigma$ upper limit of 1.2 mJy on the radio flux density of \taub. Later, \cite{Lynch2018} deduced a limit of 19.0 mJy from the survey of \cite{Lenc2018} at 200 MHz. The tentative detection of a circularly polarized signal at 14–21 MHz with flux density 890 mJy by \cite{Turner2021} has not been replicated. Most recently, \cite{Turner2024} placed an upper limit of 165 mJy at 15–39 MHz, derived from the non-detection in their follow-up observations with \mbox{LOFAR}. Further, \mbox{tau Boo} system was also probed in a higher frequency domain, where \cite{Stroe2012} placed an upper limit of 0.13 mJy at 1.7 GHz using their observations with the Westerbork Synthesis Radio Telescope (WSRT). However, our results predict a maximum emission frequency of 48 MHz. Therefore, these upper limits are irrelevant except for the recent limits by \cite{Turner2021} and \cite{Turner2024}. Our predicted flux density is within the upper bounds of the relevant observations, meaning there is a need for instruments of higher sensitivity to detect emissions from \taub{}. One exciting possibility here is to consider the addition of NenuFAR's antennas to those of LOFAR's, which almost doubles the sensitivity of the LBA \citep{Zarka2015b}. Indeed, our calculations of the sensitivity of NenuFAR agree, and \taub{} is found to be observable.  However, following their non-detection and using 53 of the 96 total mini-arrays of NenuFAR, \cite{Turner2023} had been able to place an upper limit of only 1.5 Jy and 590 mJy at 14.8–52.1 MHz range, assuming the sensitivity of NenuFAR is that of LOFAR and twice that, respectively. 

\subsection{Implications for Demographics}
\begin{figure}
    \centering
    \includegraphics[width=0.475\textwidth]{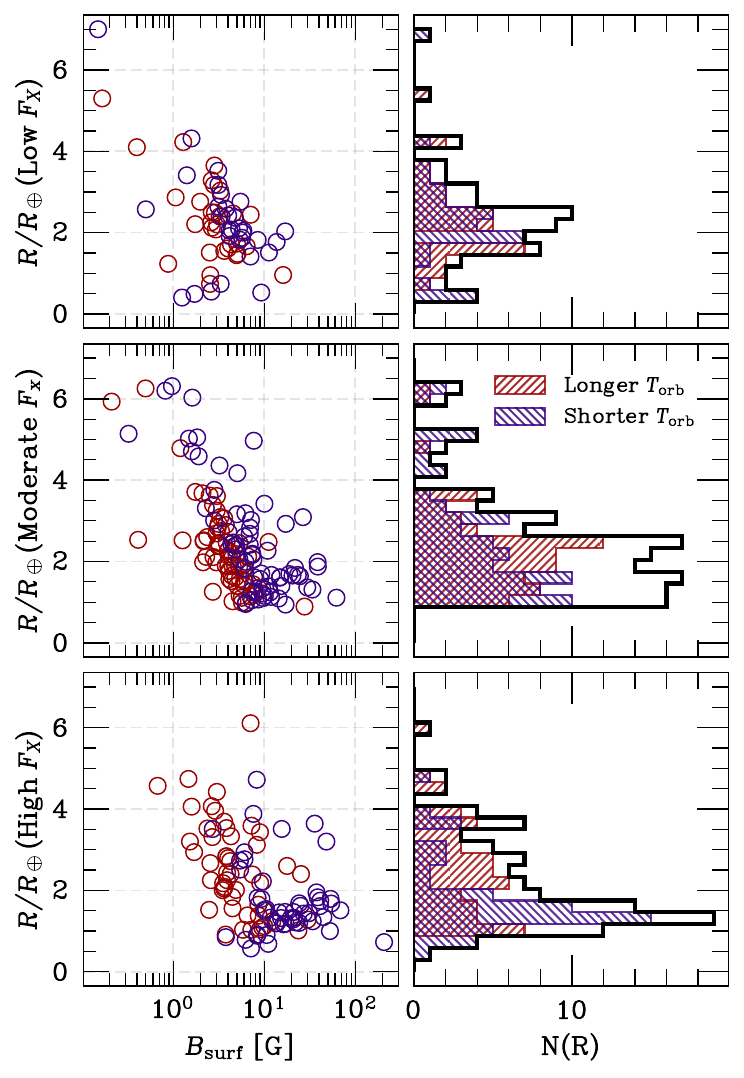}
    \caption{The radius and magnetic field strength distributions of the close-in ($a < 0.1\,\mt{AU}$) exoplanets in the radius range 1–7 $R_\oplus$. Three subsets are formed on the amount of X-ray flux incident on the planets: Low ($ <10\, F_{X,\oplus}$), Moderate (10–100\,$ F_{X,\oplus}$), and High ($ >100\,F_{X,\oplus}$), where $F_{X, \oplus}$ is the X-ray flux incident on Earth due to irradiance from the Sun.  Each subset is divided into two groups from the median orbital period, with the color blue marking the shorter-period group and red marking the longer-period one.}
    \label{fig: valley}
\end{figure}

As discussed in Section~\ref{section: Introduction}, according to the photoevaporation model of atmospheric escape, planets with weak magnetic dynamos are stripped of their thin envelopes, causing them to shrink in radius. An interpretation is that the smaller planets ($R < 1.5\,R_\oplus$) are more likely to possess weak magnetism. However, this is not taken into account in our model, specifically in Equation~\ref{eq: moment}, where the planetary magnetic moments are scaled with four basic parameters. In fact, due to the variations in $\rho_c$ being of order unity, and $\sigma_c$ is taken to be a constant among subsets of exoplanets with similar radii (and thus similar $r_c$), $\mu$ is dominantly controlled by the rotation rate. Since our model assumes tidal synchronization for close-in planets ($a < 0.1\,\mt{AU}$), the \textit{orbital} periods of the planets determine their magnetic momenta and shorter periods result in stronger fields. This fact is reflected at the bottom right corners of Figure~\ref{fig: valley}'s left panels, where the radii and the surface magnetic field strengths of the small, close-in planets in our sample are visualized.
The sample is analyzed in three groups with different conditions of incident X-ray flux, the main driver of photoevaporation. The stellar X-ray fluxes were estimated using the scaling relation of \cite{Ayres1997} mentioned in Section~\ref{subsection: wind model} and scaled to the orbital distances of planets. Moreover, each flux group is divided into two populations of equal size by the length of the planetary orbital periods. The shorter-period planets, marked by the color blue, are seen to accumulate in the rightmost parts of the panels due to their high $B_\mt{surf}$ values. 

Furthermore, we note that bodies with the lowest magnetic fields are found near the less violent stars. Finally, the radius valley around $2\, R_\oplus$ and the lack of short-period Neptune-sized planets are illustrated in the right panel of the Moderate flux group in Figure~\ref{fig: valley}. As expected, the distribution of the shorter-orbital-period population peaks at smaller radii compared to that of the longer-period population, resulting in the overall bimodal distribution of planetary radii. In the lower and higher flux groups, the radius distributions peak around values higher and lower than $2\, R_\oplus$, respectively, as expected from the evaporation model.

Although the photoevaporation model does not favor small, close-in exoplanets with high magnetic field strengths, our model does. In fact, except \added{\mbox{GJ 674 b}, \mbox{V1298 Tau c}, and \mbox{GJ 876 d}}, all sub-Jovian candidates listed in Table~\ref{tab: obs} belong to this category, spanning a radius range of 1.5–1.9$\,R_\oplus$. Future non-detections of auroral radio emissions from these targets could be explained by their lack of the predicted magnetic fields in accordance with the photoevaporation model. On the other hand, a significant number of detections from these targets could challenge the photoevaporation explanation for the demographics of the relevant exoplanets. However, given the lack of any robust magnetic field detections, such hypothesis tests remain inaccessible.

Moreover, the lack of atmospheres of small, close-in rocky worlds could be due not only to photoevaporation but also to core-powered mass loss \citep{Ginzburg2018, Gupta2019} or gas-poor formation \citep{Lee2014, Lee2016}. While atmospheric loss due to photoevaporation might imply a weak magnetosphere that fails to counter the loss of the gaseous envelopes, such a loss could also occur in the presence of a robust magnetic field. In particular, CMEs might be a key driver of atmospheric loss, compressing the magnetosphere of the planet and exposing its atmosphere directly to CME plasma flow \citep{Khodachenko2007, Kay2016}. Combined with the episodic extreme irradiation events due to stellar flares, close-in exoplanets may have eroded atmospheres despite being magnetized. In this regard, magnetospheric constraints on such exoplanets through \mbox{(non-)detections} of auroral radio emissions will be critical. However, the necessity of analyzing planetary formation history should also be noted to evaluate core‐powered mass loss and gas‐poor formation scenarios.

\subsection{Theoretical Limitations}
We now turn to the weaknesses and the limitations of the model developed above. First, several factors that might prevent the detection of such emissions have been omitted, and there is room for theoretical improvement. \cite{Nichols2016} calculated the auroral radio powers of exoplanets while accounting for magnetospheric convection saturation and found that HJs are strongly inhibited from fully dissipating away the incident Poynting flux on them, implying the magnetic RBL predictions are overestimates. Further, \cite{Weber2017} and \cite{Daley-Yates2018} reported that the extended ionospheres associated with the expanding atmospheres of HJs might possess plasma of much higher frequency than the cyclotron frequency, and thus the CMI-driven emissions could remain trapped in the local plasma of the planet's ionosphere. Elaborating, \cite{Weber2018} discovered that supermassive HJs such as \taub{}, thanks to their strong gravitation, have exobases closer to their surface, allowing a depleted plasma region to form between the exobase and the magnetopause. This favors supermassive HJs, such as the HJs we have as our opportune targets, with radio emissions that can penetrate their local environment. 

Further, although we filtered the exoplanet sample by the ability of their radio emissions to escape the local stellar wind plasma at the semi-major axis of the orbit (Equation~\ref{eq: plasmafreq}), such a requirement may not always be restrictive enough for observations, as mentioned in Section~\ref{subsection: Opportune Targets}. Through the orbital motion of a planet, due to different geometrical configurations explored in \cite{Kavanagh2020}, its emission might need to propagate through denser regions of the stellar wind and from inside the radio photosphere, effectively being absorbed or eclipsed through the so-called free-free absorption processes. Since most known exoplanets are transiting their stars, they are particularly prone to such eclipses, given their orbital inclinations. Moreover, as attempted by \cite{Ashtari2022}, the directivity of the emission should be considered together with the emission latitudes and orbital geometry. For $i\sim 90\degree$, which is the case for transiting exoplanets, the specific shape of the emission cone would be critical to assess visibility from Earth. These considerations significantly reduce the observation probability of CMI driven emissions from exoplanets, and must be accounted for in forms of case studies, they are out of the scope of this work. Evidently, further theoretical work is necessary to more accurately determine the top targets for auroral emissions. This also includes detailed analyses of the geometry of the candidate systems, such as the one by \cite{Kavanagh2023} that further analyzes the geometry of induced CMI emissions.

\subsection{Outlook}
\added{As a workaround for the limitations caused by the geometry of candidate systems, it is worth noting that the advances in the radial velocity technique have extended the range of orbital inclinations for which discoveries are possible, particularly in the solar neighborhood \citep[e.g.,][]{Lockwood2014, Feng2022}. In the future, with more motivated searches, especially around Habitable Worlds Observatory target stars, an increased population of lower-inclination exoplanets is expected. Such a subset of exoplanets could provide more favorable conditions for evading free-free absorption and observability limitations imposed by the emission directivity, which could lead to higher-yield searches for magnetism-probing radio signals.}

Despite decades of investigations of exoplanetary CMI emissions, the first detection of radio emission from an exoplanet is yet to be made. New low-frequency radio telescopes such as the Square Kilometre Array \citep[SKA; ][]{Dewdney2009} currently being built in Australia will be instrumental in probing the long-wavelength regime at unprecedented sensitivity. Furthermore, space-based interferometry provides another opportunity to go even lower in frequency, opening up possibilities to detect candidates such as the ones in Table~\ref{tab: space-obs2}, with the ionosphere out of the way. For example, The Great Observatory for Long Wavelengths \citep[GO-LoW; ][]{Knapp2024} is a space-based array proposal curated for detecting magnetic fields of terrestrial exoplanets emitting in the 300$\,\mt{kHz}$–15$\,\mt{MHz}$ range, designed to be placed in an Earth-Sun Lagrange point. In addition, radio astronomy from the Moon presents exciting opportunities, as the Lunar Surface Electromagnetics Explorer \citep[LuSEE-Night;][]{Bale2023} is scheduled for launch around early 2026 to probe the low frequency ($< 50\; \mt{MHz}$) radio sky from the far side of the moon without interference from other radio sources such as the Sun, Earth, and various Lunar experiments.  With the prospective Stokes V measurements with such new and existing facilities, the circularly polarized radiation from exoplanets is expected to be eventually detected.

\section{Conclusions}
\label{section: Conclusion}
We presented the results of a computational framework we developed to predict the radio emission characteristics of $\added{1579}$ exoplanets. The CMI-driven emissions were modeled by the Radiometric Bode's Law (RBL), which relates the input magnetic and kinetic power incident on a planet to its output radio emission power. The theoretical model to determine the inputs of the RBL was integrated from the literature, as often done in such predictive research. To assess the ground-observability of the radio emission candidates favored by our model, we considered four low-frequency radio telescopes, LOFAR, NenuFAR, uGMRT and MWA, and evaluated the observation prospects of our candidates. We have found \added{16} sources with promising emission characteristics that can be detected with current ground-based facilities. They are detailed in Table~\ref{tab: obs}. Further, we presented in Table~\ref{tab: space-obs2} the five sources expected to be the brightest ones in the 0.1–10 MHz frequency range, whose CMI emissions might potentially be detected using future space-based facilities. 

All of our \added{16} top-ranked exoplanets for ground-based observations are close-in ($< 0.1\, \mt{AU}$), and \added{nine} are small with radii $R < 2\,R_\oplus$, which likely have thin or no atmospheres due to either photoevaporation, core-powered mass loss or gas-poor formation. Emphasizing the possibility of atmospheric escape due to photoevaporation and core-powered mass loss even within the presence of strong magnetic fields, we encourage observational radio campaigns targeting our smaller candidates.

Although \added{four} of our top-ranked targets, \taub{}, \ffc{}, \wb{},\added{ and \mbox{GJ 876 d},} have been previously observed with ground-based radio telescopes, the placed upper limits are not in conflict with our predictions due to several reasons analyzed in Section~\ref{subsection: comparison}. Therefore, more sensitive observations at long wavelengths are promoted. With the theoretical anticipations aligned and more sensitive observations made possible by future facilities, the likelihood of detections from targeted observation campaigns is expected to increase.

The search for exoplanetary radio signals is a noble endeavor aiming to enhance our understanding of potentially habitable worlds. Predictive studies such as this one will be key and pave the way toward the first breakthrough detections. 

\section*{Acknowledgments}
We acknowledge support from the McDonnell Center for the Space Sciences at Washington University in St. Louis. \added{We thank the anonymous referees for valuable comments and feedback on our work.} This research has made use of the NASA Exoplanet Archive, which is operated by the California Institute of Technology, under contract with the National Aeronautics and Space Administration under the Exoplanet Exploration Program.

\software{Python 3.10}
\facility{Exoplanet Archive}

%\appendix
\bibliographystyle{aasjournalv7}
\bibliography{references, export, manual}

\end{document}